\documentclass{nature}
\usepackage{hyperref}
\usepackage{epsfig}
\usepackage{color}
\usepackage[T1]{fontenc}
\usepackage{mathtools}
\usepackage{setspace}
\usepackage{graphicx}
\usepackage{caption}
\usepackage{subcaption}
\usepackage{tocloft}
\usepackage{etoc}       
\usepackage{amsfonts}
\usepackage{amssymb}
\usepackage{amsmath}
\usepackage{bm}

\usepackage[super,comma,sort&compress]{natbib}

\usepackage{hyperref}
\usepackage{verbatim}
\usepackage{float}
\usepackage{placeins}
\usepackage{graphicx}

\def\mearth{{\rm\,M_\oplus}}

\newcommand{\beginextendeddata}{%
        \setcounter{table}{0}
        \renewcommand{\thetable}{Extended Data \arabic{table}}%
        \setcounter{figure}{0}
        \renewcommand{\thefigure}{\arabic{figure}}%
        \renewcommand{\figurename}{Extended Data Figure}%
     }

\newcommand{\beginsupplement}{%
        \setcounter{table}{0}
        \renewcommand{\thetable}{Supplementary \arabic{table}}%
        \setcounter{figure}{0}
        \renewcommand{\thefigure}{\arabic{figure}}%
        \renewcommand{\figurename}{Supplementary Figure}%
     }

\title{Very wide-orbit planets from dynamical instabilities during the stellar birth cluster phase}

\author{André Izidoro,$^{1}$\thanks{Corresponding Author: \href{mailto:izidoro@rice.edu}{izidoro@rice.edu}}~~~Sean N. Raymond,$^{2}$ Nathan A. Kaib,$^{3}$ Alessandro Morbidelli,$^{4}$ and Andrea Isella$^{5}$}

\begin{document}

\maketitle

\begin{affiliations}
\item Department of Earth, Environmental and Planetary Sciences, 6100 Main MS 126,  Rice University, Houston, TX 77005, USA
\item Laboratoire d'Astrophysique de Bordeaux, Univ. Bordeaux, CNRS, B18N, all{\'e}e Geoffroy Saint-Hilaire, 33615 Pessac, France
\item Planetary Science Institute, 1700 E. Fort Lowell, Suite 106, Tucson, AZ 85719, USA
\item Collège de France, 11 Pl. Marcelin Berthelot, 75231 Paris, France
\item Department of Physics and Astronomy, 6100 Main MS-550, Rice University, Houston, TX 77005, USA
\end{affiliations}

\begin{abstract}

Gas giant planets have been detected on eccentric orbits several hundreds of astronomical units in size around other stars. It has been proposed that even the Sun hosts a wide-orbit planet of 5-10 Earth masses, often called Planet Nine, which influences the dynamics of distant Trans-Neptunian objects. However, the formation mechanism of such planets remains uncertain. Here we use numerical simulations to show that very wide-orbit planets are a natural byproduct of dynamical instabilities that occur in planetary systems while their host stars are still embedded in natal stellar clusters. A planet is first brought to an eccentric orbit with an apoastron of several hundred au by repeated gravitational scattering by other planets, then perturbations from nearby stellar flybys stabilise the orbit by decoupling the planet from the interaction with the inner system. In our Solar System, the two main events likely conducive to planetary scattering were the growth of Uranus and Neptune, and the giant planets instability. We estimate a 5-10\% likelihood of creating a very wide-orbit planet if either happened while the Sun was still in its birth cluster, rising to 40\% if both were. In our simulated exoplanetary systems, the trapping efficiency is 1-5\%. Our results imply that planets on wide, eccentric orbits occur  at least $10^{-3}$ per star.

\end{abstract}


\section*{Main Text}
The origins of planets on very wide orbits remain hard to explain. Planets orbiting a few hundred to a few thousand au from their host star are dynamically isolated from gravitational perturbations from any inner planet ($\lesssim$10~au) as well as from Galactic perturbations\cite{tremaine93,wyatt17} ($\gtrsim$10,000~au). Several hundred au is also wider than the size of most planet-forming disks\cite{baeetal23}, and the disk that formed the Solar System was smaller than $\sim 80$~au.\cite{kretke12} Yet to date, about a dozen exoplanets with orbits wider than 100 au have been discovered\cite{Bohnetal20,zhangetal21,markusetal21,bohnetal21,gaidosetal22}; given observational challenges,\cite{currieetal22} most are several times more massive than Jupiter. Some wide-orbit exoplanets may be on near-circular orbits (such as the HR 8799 system with four super-Jupiters\cite{marois10}) while others are on eccentric orbits,\cite{nguyen21,faherty21} and still others do not even orbit a star and are instead 'free-floating'.\cite{mroz17,miretroig22} It has also been proposed that Solar System hosts a yet-to-be-detected wide-orbit planet called {\em Planet Nine}.\cite{batyginbrown16,brownbatygin16,trujillosheppard14} The evidence for Planet Nine is circumstantial:  the orbits of many of the widest-orbit Trans-Neptunian Objects (TNOs)  appear to be clustered in longitude of perihelion and orbital pole
position.\cite{batyginbrown16, brownbatygin21, batyginbrown21} This clustering could be explained by a planet on an exterior orbit through a combination of resonant and secular dynamics~\citep{batyginmorbidelli17,batygin19}.  An alternative possibility is that the observed alignments are artifacts from observational biases combined with mere chance.\cite{shankmanetal17} However, it remains unclear whether any set of biases can explain the clustering of this particular set of distant TNOs.\cite{batygin19} 

Dynamical instability is likely to play a central role in the production of wide-orbit planets, as it is believed to be ubiquitous in giant planet systems\cite{raymond20} and places planets on wide (albeit unstable) orbits. There is evidence that the Solar System's giant planets underwent two dynamical instability phases. First, as $\sim 5 \mearth$-scale planetary embryos migrated, collided and grew into the ice giants, many embryos were ejected after close encounters with Jupiter and Saturn.\cite{izidoroetal15c,helledetal20} Second, during the giant planets' final dynamical instability, 1-2 additional ice giants were ejected.\cite{nesvornymorbidelli12} The broad eccentricity distribution of observed giant exoplanets\cite{butler06,udrysantos07} is thought to be the result of dynamical instabilities in 75-95\% of systems.\cite{juric08,chatterjee08,raymond10} Most dynamical instabilities are likely to take place during or shortly after the main phase of planet formation. The detection of $\sim100$ free-floating planets in a single young stellar cluster\cite{miretroig22} is strong circumstantial evidence that many dynamical instabilities occur early. In the Solar System, the icy embryo instability  (i.e. during the formation of Uranus and Neptune)  happened during the Sun's gaseous disk phase, but the timing of the final instability is uncertain but constrained to have happened within the first 10-100 million years of Solar System history.\cite{Nesvornyetal18b,liuetal22} Yet stars do not form in isolation, but rather in groups containing hundreds to thousands of members called gas-embedded clusters (hereafter refereed to as ``embedded cluster'').\cite{ladalada03,adamsetal10,pfalzner13,daleetal15,adamoetal20}  During the $\lesssim$10~Myr embedded cluster phase  -- the lifetime for a typical cluster -- stars undergo close mutual encounters\cite{adamslaughlin01,portegiesetal10,malmberg11,parkeretal12} and in some cases a planet can be captured from another star\cite{laughlinadams98,peretsKouwenhovenetal12,liadams16,mustilletal16,parkeretal17,vanelterenetal19,daffern-powell22}. The probability of capture increases if the planet's orbital distance from its host star is sufficiently large, as this enhances the likelihood of being captured by passing stars.\cite{liadams16,mustilletal16,parkeretal17}, with typical capture probabilities of the order of $\lesssim$1\%. Alternate mechanisms for the formation of wide-orbit planets include formation by gravitational collapse\cite{boss97}  of a portion of the protoplanetary disk, in-situ accretion within a very massive, distant ring of planetesimals\cite{kenyonbromley15}, and scattering from the giant planet region followed by orbital circularization through dynamical friction with an extended cold planetesimal belt\cite{erikssonetal18} or by gas dynamical friction\cite{bromleykenyon16}. However, sufficiently extended disks of gas, dust, and/or planetesimals appear to be extremely rare according to the demographics of planet-forming disks observed with ALMA\cite{baeetal23}, Gemini Planet Imager (GPI) Exoplanet Survey\cite{espositoetal20} and SPHERE.\cite{beuzit19} It remains uncertain whether the observed occurrence rate of wide disks is sufficient to account for the observed frequency of wide-orbit exoplanets.

\begin{figure*}
\centering
\includegraphics[width=.9\linewidth]{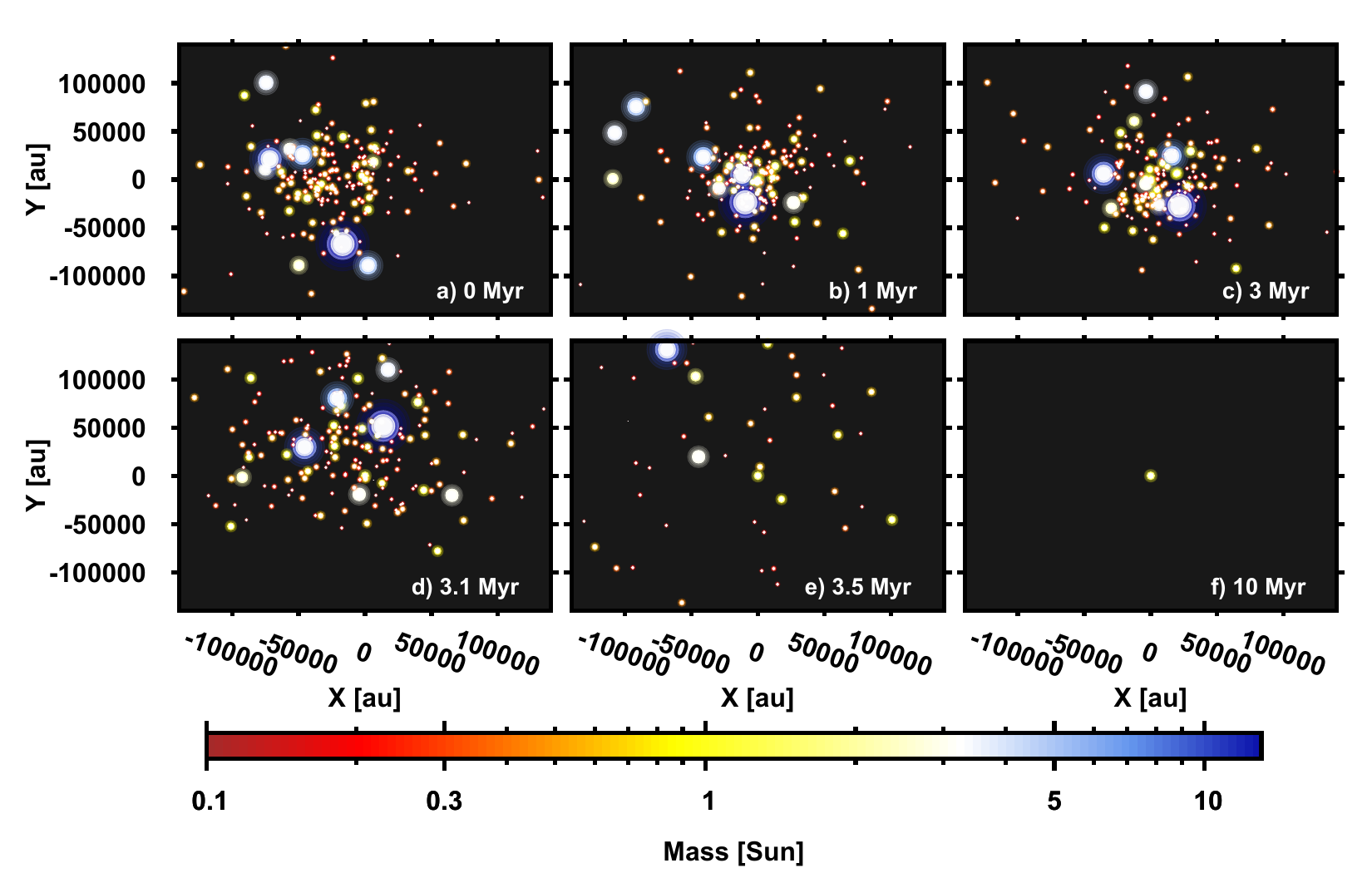}
\caption{{\bf Snapshots of the dynamical evolution of a stellar cluster with 200 stars.} Panels show the top view projection of the cluster at different times. Color-coded circles represent individual stars and the size of the circle scales with $M^{0.5}$ (for presentation purposes only), where $M$ is the star mass. In this nominal simulation, a solar-mass star is at the center of the reference frame. We refer to this star as the host star. We model planet formation and evolution only around the host star (or binary star system; see Methods). Stars in the cluster gravitationally interact with each other, with the host star, and its planets. All stars and planets tidally interact with the gas component of the cluster, which is modeled following the Plummer model, by setting the Plummer radius as $R_{\rm c}$= 40,000 au (see Methods). The gas cluster’s density is assumed constant over 3 Myr (panels a, b and c). Individual stars may  eventually be ejected from the cluster during the gas cluster phase due to star-star encounters, but most stars remain bounded to the cluster until the gas of the cluster dissipates. In this simulation, this happens at 3 Myr, and stars start to disperse (panels d and e ). At 10 Myr – the end of the simulation – stars in the cluster have already dispersed significantly and moved away from the host star (panel f).}
\label{fig:clusterevol}
\end{figure*}

Here we show that planetary dynamical instabilities during the embedded star cluster phase naturally produce wide-orbit planets, which we broadly define as having semimajor axes between 100 and 10,000 au and being dynamically detached from a closer-in set of planets, with periastron distances large enough to avoid interaction with other planets. We performed dynamical simulations of different planetary systems embedded in star clusters with a range of properties.  We simulated the following planetary systems: both of the early Solar System's giant planet instabilities, systems of gas giant exoplanets, systems of ice giant exoplanets, and planetary systems orbiting close binary stars (so-called circumbinary systems\cite{welsh18}). Each planetary system was placed in orbit around a Sun-like star embedded in a cluster with $N_{\star}=200$ to $5000$ stars (see Fig.~\ref{fig:clusterevol}), and the full system was evolved for the stellar cluster lifetime and beyond.  We also varied the stellar density of the cluster by adjusting the cluster's characteristic size (its Plummer radius) and lifetime (See Methods) considering scenarios broadly consistent with observations and previous studies \cite[e.g.][]{proszkowadams09}. Observations show that most stars remain gravitationally bound to their natal cluster for $\sim$3-10~Myr.\cite{ladalada03,allenetal07,falletal09} During this phase, stars gravitationally interact with each other, with the remaining gas of the cloud, and with their circumstellar disks.  Close encounters among stars may sculpt and truncate protoplanetary disks\cite{wijnenetal17,cuelloetal23} (which may also be evaporated by radiation from high-mass stars\cite{winter18}), but the large observed occurrence of exoplanets\cite{fressinetal13} and disks\cite{haischetal} strongly suggest that most stars preserved their disk during this phase.  In our model we ignore the effects of stellar evolution and the presence of primordial binaries (see Methods).

\begin{figure*}
\centering
\vspace{-0.5cm}
\includegraphics[width=1.\linewidth]{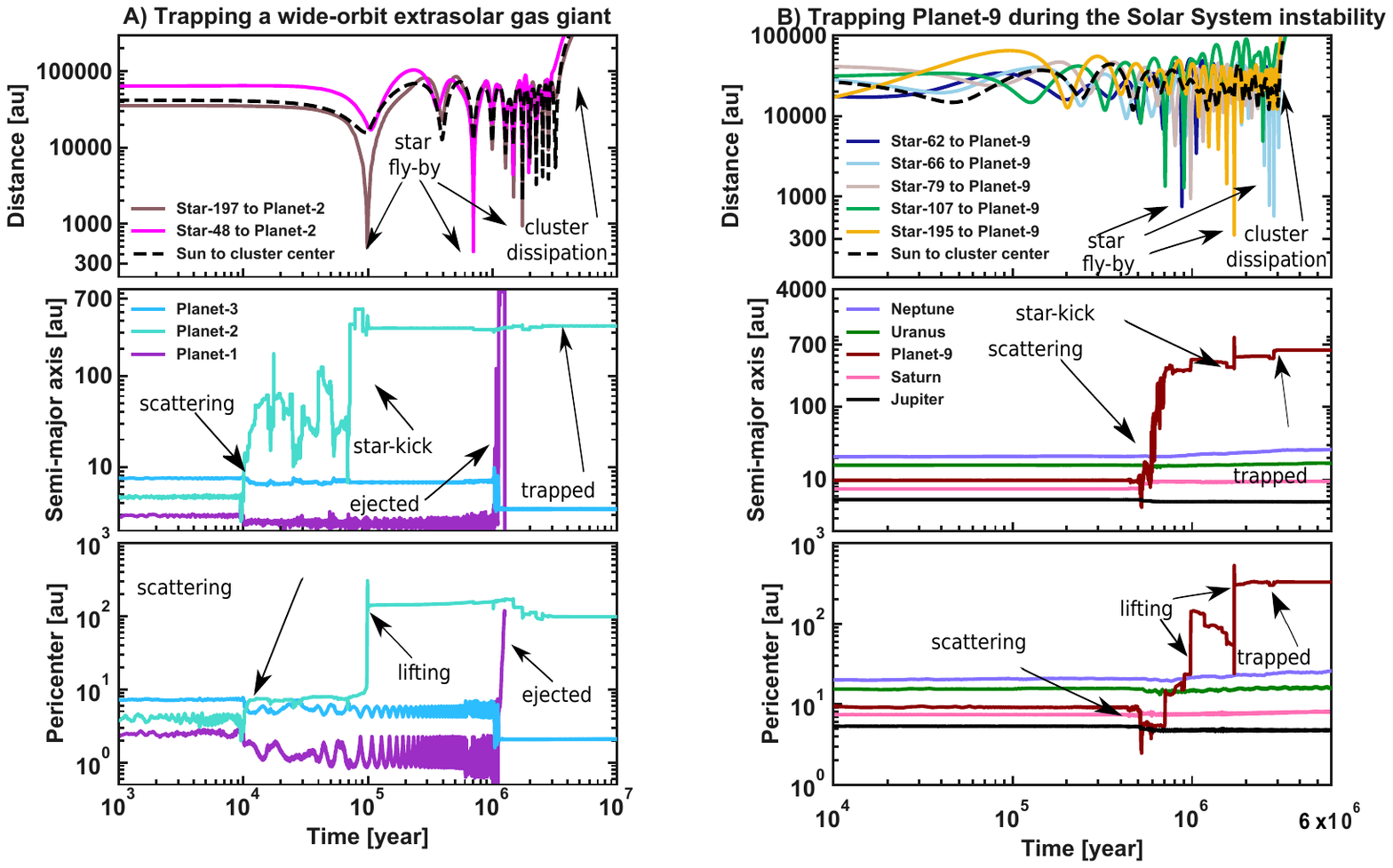}
\vspace{-1.5cm}
\caption{\footnotesize{\bf Trapping of wide-orbit planets in a extrasolar gas-giant instability  simulation and a Solar System early dynamical instability simulation.}  Both simulations of panels A and B start with fully formed planets.  The cluster contains 200 stars and $R_{\rm c} = {\rm 40,000}$~au. The host star is a solar-mass star in both simulations. {\bf A):} Trapping of a wide-orbit exoplanet. The top panel shows the distance of two specific star members of the cluster to Planet-2, as the cluster evolves. The black dashed line shows the distance of the host star to the barycenter of the star cluster. The middle and bottoms panels show that Planet-2 is scattered into a wide orbit during a planetary instability that takes place at about 10 kyr. At 100 kyr, Star-197 (brown line) flies-by and kicks Planet-2, lifting its pericenter, trapping it on a stable orbit. Star-48 approaches Planet-2 at about 0.7 Myr but the planet orbit is only weakly affected by the approach. Planet-1 is ejected from the host star at 1 Myr and Planet-3 remains on a 2.5 au-wide orbit (middle and bottom panels). The cluster dissipates at 3 Myr, as shown in the top panel. Planet-2 remains on a stable wide-orbit until the simulation is stopped at 10 Myr (see Supplementary Movie 1 showing the cluster dynamics and the approach of Star-197 kicking Planet 2 at 100 Myr). {\bf B)}: Trapping of the hypothetical Planet-9 during an early Solar System dynamical instability. Planet-9 is scattered at about 0.4~Myr onto a wide orbit after having close-encounters with Jupiter. Stars 107 and 62 fly-by and kick Planet-9 at 0.7~Myr and 0.9~Myr, respectively, slightly lifting its pericenter. At about 2~Myr, a very strong encounter between Planet-9 and star 195 dramatically changes the orbit of Planet-9, further lifting its pericenter. The cluster dissipates at 3 Myr and the simulation stops at 6 Myr. Orbital elements are given with respect to the barycenter of the host star system. At the end of the simulation, the orbits of the four gas giant planets are qualitatively equivalent to those of the real planets (see Supplementary Movie 2).}
\label{fig:trapping}
\end{figure*}

A planet may be trapped on a wide orbit after being scattered onto a highly-eccentric orbit with a semimajor axis of hundreds of au, followed by an external perturbation from either a star passing within $\sim 1000$~au, a rapid change in the external gravitational potential\cite{brasseretal06} as the host star passes near the cluster center,  or encounters with a dense filament of the gas cluster.  If the external perturbation provides a gravitational kick in the right direction (leading to an increase in orbital angular momentum), the scattered planet's eccentricity decreases and the corresponding increase in its periastron distance protects it from further planetary scattering. Fig.~\ref{fig:trapping} show  examples of successful trapping of wide-orbit planets in two different types of planetary systems. Fig.~\ref{fig:trapping}A) show one system of gas giant exoplanets and  Fig.~\ref{fig:trapping}B) show a  solar system giant planet dynamical instability. Our simulations modelling dynamical instabilities of extrasolar gas giant planets are designed to   broadly match the eccentricity distribution of observed radial velocity planets\cite{rasioford96,raymond10,beuagenesvorny12} (see Extended Data Figure 1). Our solar system dynamical instability simulations setup is inspired by the most successful configurations in explaining properties of the outer solar system~\cite{levisonetal11,nesvornymorbidelli12,deiennoetal18}. Fig.~\ref{fig:trapping} shows that, once trapped,  wide-orbit planets remain vulnerable to external perturbations for the remainder of the cluster phase (as in Extended Data Figure~2), but once the cluster dissipates the wide-orbit planet is usually stable over a long term.

\begin{figure*}
\includegraphics[width=.95\linewidth]{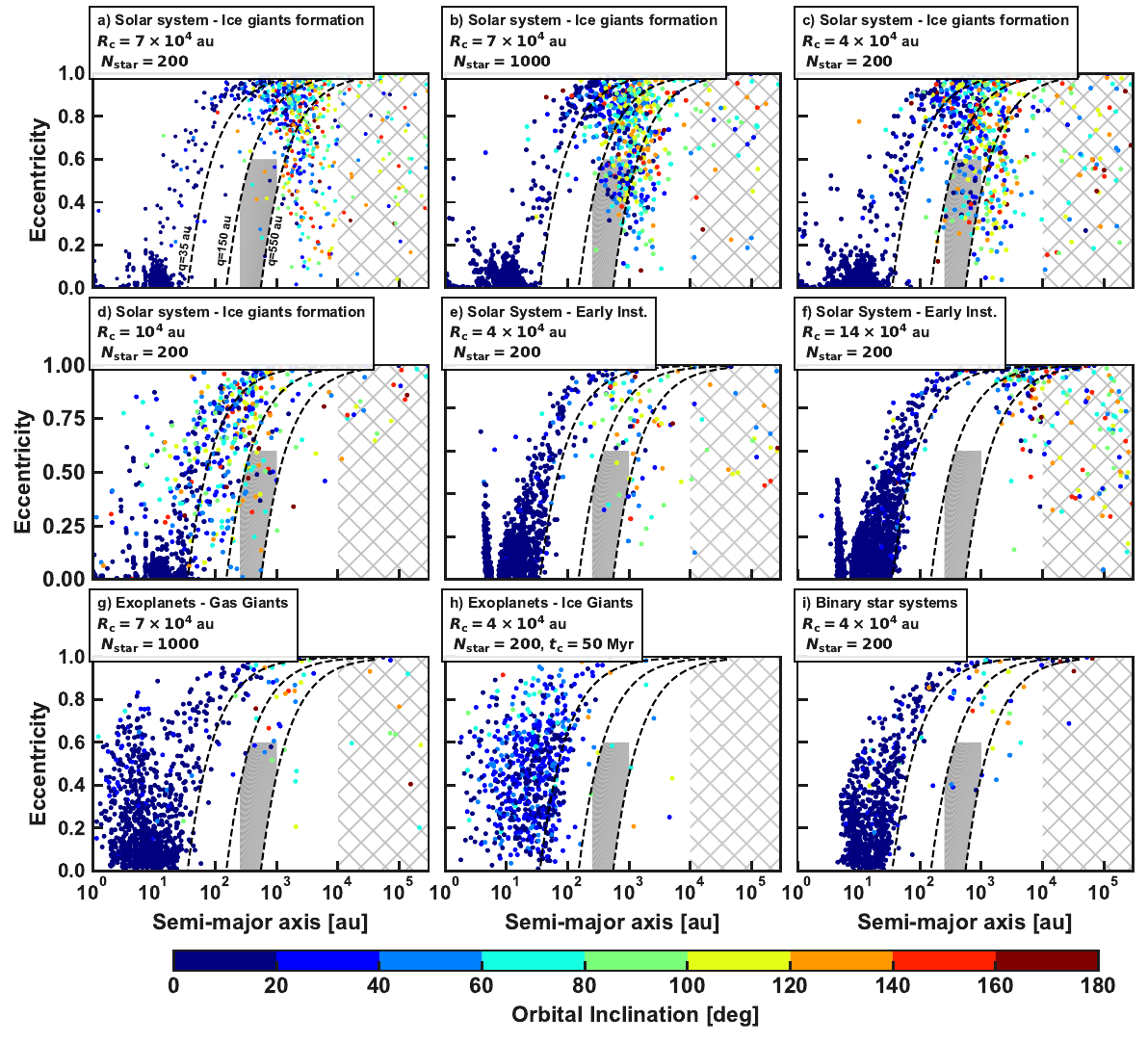}
\caption{\footnotesize{\bf Final orbital distribution of planets produced in early dynamical planetary instabilities taking place in different planetary systems embedded in stellar clusters with different configurations}. $N_{\rm star}$ represents the number of stars in the cluster. $R_{\rm c}$ is the cluster Plummer radius, $t_{\rm c}$ is the timing of cluster dispersal. Panels a, b, c, and d show the orbital distribution of planets in simulations modeling the accretion of Uranus and Neptune.\cite{izidoroetal15c} Panels e) and f) show the distribution of planets in simulations modeling the Solar System's final instability.\cite{clementetal18,ribeiro20,liuetal22} Solar System panels show all our simulations, including cases that failed to broadly reproduce the solar system architecture at the end. The three black dashed curves mark the locations where orbital pericenters are 35, 150, and 550 au. The gray hatched area corresponds to the region ($>10000$~au) where planets are likely to be affected by the Galactic tide and passing stars, implying that their orbits may evolve over Gyr timescales.\cite{higuchietal07} The gray filled area shows the expected location of Planet-9  ($250<a<1000$~au; $e<0.6$; $150<q<550$ au), roughly within 3$\sigma$ uncertainties\cite{batygin19,brownbatygin21}, and it is showed only for reference in those cases not corresponding to the Solar System. Panel g shows the distribution of planets in extrasolar gas-giants instability simulations  modeling the dynamical evolution of three gas giant planets (see Methods). Panel h shows the outcomes of planet-planet scattering in simulations modeling the dynamical evolution of six Neptune-mass planets. In this particular case, the cluster dissipation timescale is $t_{\rm c}$=50 Myr, much longer than our nominal values of 3-5 Myr. Panel i) shows the final distribution of planets in simulations modeling the dynamical evolution of gas giant planets around close binary stars (circumbinary planets).}
\label{fig:trapped_orbits}
\end{figure*}

The demographics of trapped wide orbit planets depend both on the unstable planetary systems themselves and on the cluster properties (Fig.~\ref{fig:trapped_orbits}).  The strength of the external perturbations felt by scattered planets determines the minimum orbital size at which a scattered planet can be trapped.\cite{tremaine93,wyatt17}  Stronger perturbations imply a higher stellar density, from a larger number of stars or a more compact cluster. For instance, Figures ~\ref{fig:trapped_orbits}-b and c produced very similar distributions of trapped planets between $\sim$400 and $\sim$1000 au. Although the number of stars and cluster sizes are different in these two cases, they correspond to similar cluster densities. This result is consistent, for instance, with those of \citep{brasseretal06} who demonstrated that the final distribution of planetesimals trapped at large semi-major axis depends on the mean density that the sun encounters. We also note in Fig.~\ref{fig:trapped_orbits} that planets at large semi-major axes ($\gtrsim$ 10000~au; see hatched region) are likely to evolve under the influence of galactic tides and passing stars over Gyr timescales.\cite{higuchietal07,kaibquinn08,baileyfabrycky19,raymond23}
In addition, planets with orbital pericenters below 35 AU, as well as those with pericenters between 35 and 150 AU and exhibiting high orbital eccentricities ($\sim$0.9–1), may remain partially coupled to the inner system and may not maintain long-term stability. These planets are excluded from our quantitative analysis.

For the Solar System, only certain cluster parameters are compatible with Planet Nine. Dynamical simulations and stability arguments constrain  Planet Nine's mass  to be $4-12\mearth$, its orbital semimajor axis to roughly lie between $\sim$250 and $\sim$1000 AU, eccentricity lower than $\lesssim$0.6, with a modest inclination of $\sim$10-30$^\circ$ relative to the ecliptic plane~\citep{batygin19, batyginbrown21}.  These values are broadly consistent with 3$\sigma$ uncertainties from best fit-solutions derived in \cite{batyginbrown21}. The parameter space for Planet Nine is mostly empty in simulations of ice giant formation embedded within a cluster with relatively weak external perturbations (Fig.~\ref{fig:trapped_orbits}a). In that case, most trapped wide-orbit planets have much larger semimajor axes than the current constraints on Planet Nine.\cite{batygin19,brownbatygin21} In contrast, a cluster with more stars or a smaller cluster size (Fig.~\ref{fig:trapped_orbits}b and Fig.~\ref{fig:trapped_orbits}c) can readily trap scattered planets in the appropriate orbital parameter space (see Extended Data Figure 3 for two examples of simulations producing wide-orbit planets during the accretion of Uranus and Neptune). For a very packed cluster (Fig.~\ref{fig:trapped_orbits}d), external perturbations are so strong that most trapped planets have orbits with semimajor axes much smaller than the Planet Nine constraints.  The same trends in the orbits of trapped wide-orbit planets are seen among all sets of simulations, although the trapping efficiency varies significantly depending on the planetary setup.

\begin{figure*}
\vspace{-1cm}
\includegraphics[width=1\linewidth]{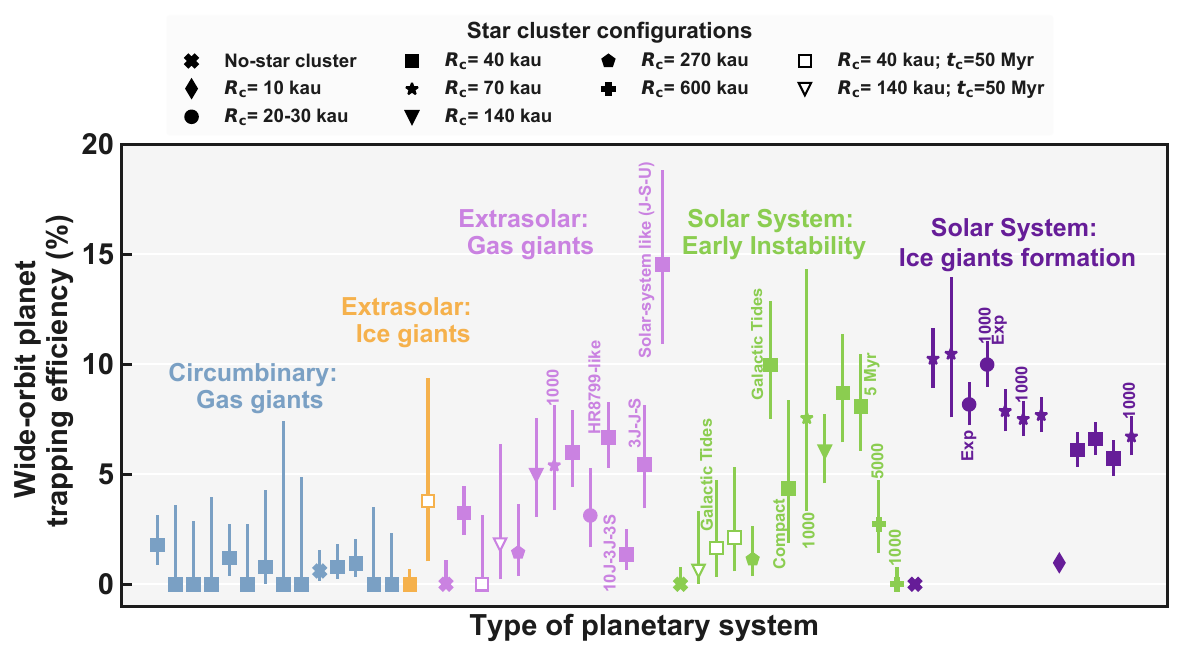}
\caption{\footnotesize{\bf Wide-orbit planet trapping efficiency in different planetary systems embedded in star clusters with different configurations.} The point shape gives the star cluster configuration as shown at the very top of the panel. Crosses are used to represent cases with no star cluster (single star or binary star systems). Points are color-coded depending on the planetary system architecture, as described in Methods. The trapping efficiency represented by each data-point is computed from at least 100 simulations, and in most cases it corresponds to 500 or 1000 simulations.  In total, this figure shows the results of approximately 30,000 numerical simulations. The error-bars are given by a binomial test considering $N_{\rm wide}$ (number of successes) and $N_{\rm wide} + N_{\rm ejec}$ (number of tries). The open-triangle and open-squares show systems with ad-hoc long-lived star clusters ($t_{\rm c}\approx50$~Myr). Data points with identical colors and shape represent initial planetary systems in the same category but with a (slightly) different planetary architecture (see Methods) or gas disk model setup. The number of stars considered in each cluster configuration is added next to the respective data point when it is larger than 200 stars (1000 or 5000 stars).  The smaller labels indicate the following: ``Galactic tides'' are systems in which the galactic tidal field was included\cite{tremaine93,wyatt17}, ``1000'' or ``5000'' are clusters with that number of stars, ``5 Myr'' are simulations with a cluster duration of 5 Myr, ``Exp'' represent cases where the gas cluster dissipates exponentially in a timescale of 0.5~Myr (rather than instantaneously at $t_c=$3 or 5~Myr), ``HR 8799-like'' are systems that started from a system akin to the HR 8799 system of super-Jupiters\cite{marois10}, ``10J-3J-J'' and ``3J-J-S'' refer to the masses of the three planets in those systems (e.g., 3J indicates $3~{\rm M}_{Jup}$), and ``Compact'' represents the case where the pre-instability solar system giant planets have more compact orbits than those in our nominal scenario.~\cite{gomesetal18}}
\label{fig:trapping_eff}
\end{figure*}

We define the trapping efficiency as the probability that a single scattered planet will be trapped on a wide stable orbit. The trapping efficiency is computed as $\epsilon$ = $N_{\rm wide}/( N_{\rm wide} + N_{\rm ejec})$, where $N_{\rm wide}$ is the final number of planets bounded to the host star with orbital pericenter $q>150$~au, $a<10000$~au, and  orbital eccentricity $e<$~0.9, and $N_{\rm ejec}$ is the final number of  planets ejected ($e>$1) from the host star.  The trapping efficiency links the abundance of wide-orbit planets with that of free-floating, or 'rogue' planets (Fig.~\ref{fig:trapping_eff}).  We found that a Solar System-like setup is optimal for trapping wide orbit planets among those that we tested, with an efficiency among different sets of simulations that is $\sim$5-10\%. Given that the Solar System is likely to have ejected multiple massive planets -- several during the ice giants' formation\cite{izidoroetal15c} and 1-2 during the final giant planet instability\cite{nesvornymorbidelli12} -- the overall likelihood of trapping a planet on a wide orbit similar to Planet Nine-like is significant. Assuming that 5 planets were ejected, each with a capture probability of 10\%, binomial statistics imply a probability of $\sim$40\% for the capture of a wide-orbit planet that could represent the putative Planet Nine.  If, instead, only three planets were ejected during the cluster embedded phase, each with a capture probability of 5\%, then the total probability drops to 14\%.

Among simulations designed to match the distribution of distant giant exoplanets,  namely planets with masses between $\sim$0.5 and $\sim$10$M_{\rm jup}$ \citep{rasioford96,chatterjee08,raymondetal10,beuagenesvorny12}, the trapping efficiency is $\sim 1-5\%$, somewhat lower than for Solar System cases.  The only exceptions are simulations in which the giant exoplanets' mass distribution was deliberately chosen to be ``Solar System-like'' and included planets with mixed masses as in the Solar System.  Exoplanets systems containing only ice giants had very low trapping efficiencies ($\lesssim 1\%$), as did circumbinary gas giant systems (Fig.~\ref{fig:trapping_eff}). Our simulations deliberately chosen to be ``Solar System-like'', including only ice giants, and those corresponding to circumbinary gas giant systems are exploratory in nature and produce orbital and mass distributions that are currently unconstrained by observations. 

 In this work, we focus on planetary dynamical instabilities taking place around solar-mass stars. The key factor in determining the trapping efficiency is the length of time during which a scattered planet remains on a wide enough orbit to interact with the external (cluster) gravitational field.  As a general rule, more massive planets tend to have shorter-lived instabilities than lower-mass planets.\cite{raymond10} Circumbinary gas giants receive additional gravitational kicks from their central (binary) stars, and scattered planets have very little time to interact with the stellar cluster  (before they are ejected). These are relatively violent dynamical instabilities where ejections happen quickly after the onset of the instability. The other extreme are systems containing only ice giants, whose instabilities evolve so slowly, and  are relatively ``weak'', that scattered planets are rarely on wide enough orbits to be captured before the cluster dissipates. This is made clear by one set of simulations in which the cluster lifetime was deliberately extended to 50 Myr and the trapping efficiency increased five-fold (Fig.~\ref{fig:trapping_eff}).  The sweet spot  within the scenarios explored in our simulations appears to be systems like the Solar System that have a mixture of different masses, including both ice giants and gas giants.  The optimal setup that we tested is the formation of the ice giants, as this involved a series of  ``moderate strength and duration'' instabilities rather than a single relatively violent one (as in the case of giant exoplanet instabilities and those of circumbinary gas giants) or the overly weak instabilities associated with scattering among multiple ice giants.

In the ice giant formation scenario, each moderate strength instability acts to scatter one or two planetary embryos. One may hypothesize that the gaseous disk, by damping the dynamical excitation of gas giants and facilitating the growth of ice giants, could reduce the intensity of scattering events, thereby prolonging the phase of interactions with passing stars. However, its overall influence appears to be relatively modest. Instead, the dominant factors seem to be the masses and the number of planets in the system. This becomes particularly clear when we compare the trapping efficiencies in the J-S-U case shown in Figure \ref{fig:trapping_eff} --which excludes the effects of the gaseous disk-- with those from our nominal ice giant formation simulations. In a more general picture, the sweet spot seems to correspond to scenarios where the planet-planet scattering phase -- characterized by an ``optimal planets-mass and number of planets'' -- occurs on a timescale comparable to the cluster's lifetime.  Note that the planet-to-star mass ratio and the location where instabilities take place in a planetary system may also influence the strength and outcome of dynamical instabilities (e.g. for instance,  scattering in the innermost regions may favor collisions instead of long-scattering phases\citep{izidoroetal17}), which, in turn, may reflect on trapping efficiencies. While the planet-to-star mass ratio determines how strongly the planet is bound to its host star, the interaction between the host star and the passing star during a flyby -- and its resulting influence of the passing star on the planet itself -- depends on the relative masses of the two involved stars\citep{carterstamatellos23}. Interestingly, numerical simulations that account for the effects of passing stars have found no significant difference in the ejection rate of planets around high-mass and low-mass stars\citep{carterstamatellos23}.

A significant population of free-floating planets has been detected, both by direct imaging in young stellar clusters\cite{miretroig22} and by microlensing\cite{mroz17,mroz20,clanton17}.  Many were likely scattered out from their home systems in the same dynamical instabilities responsible for the eccentric orbits of giant exoplanets.\cite{veras12}  Our simulations imply the existence of a significant population of planets trapped on wide orbits, that is connected to the free-floating planet population by the trapping efficiency from Fig.~\ref{fig:trapping_eff}.  Exoplanet surveys have shown that $\sim$1-10\% of all main sequence stars host giant exoplanets, integrated over all stellar types\cite{fultonetal21,bowler16,suzuki16}. To match the observed eccentricity distribution,\cite{butler06,udrysantos07} roughly 75-90\% of giant exoplanet systems must be the survivors of dynamical instability.\cite{juric08,chatterjee08,raymond10} The number of giant planets ejected per instability depends on the number of gas giants that formed in the system,\cite{veras12} as well as the number of ``innocent bystander'' planets that were ejected as a consequence of the instability.\cite{raymond11}  We expect a lower limit of one planet ejected per instability, although this number could easily be an order of magnitude higher.\cite{izidoroetal15c,veras12,raymond11} Assuming a trapping efficiency of $\sim 1-10\%$ (Fig.~\ref{fig:trapping_eff}), this implies an overall occurrence of trapped wide-orbit planets on the order of $10^{-3}$ (range of plausible values: $7.5\times 10^{-5}$ to $9 \times 10^{-3}$).  Observational surveys constrain the occurrence rate of planets (0.5-13~$M_{Jup}$) at 100-1000~au\citep{bowler16,durkanetal16} to be lower than $\sim$5-9\%, and that of planets (1-13~$M_{Jup}$) at 1000-5000~au to be lower than 3\%.\cite{baronetal18} Although these upper limits are higher than our derived value, they are not inconsistent with it. It is important to note, however, that only a few existing studies have characterized the occurrence rate of planets at such wide separations, highlighting the need for further observational efforts. We suggest that wide-orbit planets  around solar-type stars should be more common around stars that host gas giants; given the strong correlations between gas giant frequency and both stellar metallicity\cite{fischer05,dawson13} and stellar mass,\cite{johnson10} we expect that the easiest place to search for wide-orbit planets is around high-metallicity stars somewhat more massive than the Sun.

We expect most trapped wide-orbit planets to survive for tens of billions of years.  Within the local galactic context, a star only passes within 1000 au of the Sun every billion years or so, statistically speaking.\cite{adamsetal10,zink20}  Planets with orbital semimajor axes of hundreds of astronomical units are mostly protected from such external perturbations after the cluster phase. They can be lost during post-main sequence evolution of their host stars\cite{veras16} or during exceptionally close stellar passages,\cite{raymond24} which may simply require an extremely long wait.\cite{zink20}

\newpage
\section*{Methods}

\subsection{Stellar Cluster}
    To model the perturbations of an embedded cluster environment on young planeary systems, we use the MERCURY integrator\cite{chambers99,chambersetal02} as it has been modified in Kaib et al\cite{kaibetal18}. This integrator contains three classes of simulation objects: planetary objects, a (wide)-binary companion, and additional stellar objects. Planetary objects are integrated about their central star in standard democratic heliocentric coordinates\cite{duncanetal98}, while the binary star is integrated on a perturbed Keplerian orbit as in Chambers et al.\cite{chambersetal02}. In our nominal version of the code, there is no wide-binary companion (associated with the host star), so this class of object is not used when we model planetary systems orbiting single stars. Meanwhile, the additional stellar objects make up the other cluster members, and they are integrated using a simple $T+V$ leapfrog scheme about the system center-of-mass\cite{kaibetal18}. 
    
    In addition to discrete stellar masses, our simulations also attempt to model the smoother potential of the gaseous component of the embedded cluster. To do this, we represent the gaseous component with a fixed background Plummer potential centered on the discrete simulation particles' center-of-mass\cite{plummer11}. This implementation is of course an approximation in numerous ways. Most notably, as a fixed background potential, the gas does not respond to the gravitational potential of stars in the cluster. However, since cluster star formation efficiencies are generally of order 10--25\%\cite{ladalada03}, and the stars' spatial distribution with respect to the cluster center-of-mass should exhibit some level spherical isotropy, we should expect the back reactions of the overall gas distribution due to the stars to be a relatively minor process within the cluster. In addition, modeling the gas potential with a Plummer potential neglects the complex, irregular substructure of real embedded cluster gas distributions \citep{allenetal07}. Nevertheless, embedded clusters generally exhibit a level of spherical symmetry and have a peak central gas density that falls away toward the cluster edge, and we use a Plummer potential to capture the tidal forces due to this embedded cluster feature. This same approach has been employed in several previous works studying cluster perturbations on planetary systems \citep{brasseretal06}. In all our simulations, we assume a star formation efficiency of 10\%.

    In our simulations, the central star (or binary; see next section where we describe how we model the circumbinary scenario), planets, and planetesimals feel the effects of the gas-cluster (Plummer potential) and those of the stars in the cluster. Cluster stars also gravitationally interact with each other, with  planets and planetesimals,  and the central star (or binary). Planetesimals do not gravitationally interact with each other, but  gravitationally interact with central star, cluster stars, and planets, and experience gas drag. Our algorithm also accounts for close-encounters between stars.
    
    In the coordinate system employed in Kaib et al.~\citep{kaibetal18}, two positions are particularly important. The first is ${\bm s}$, the planetary system's barycenter relative to the position of the central star of the planetary system. This can be written as 
    
    \begin{equation}
        \bm{s} = \frac{\sum\limits_{i=1}^{N_P}m_i \bm{X}_i}{m_A + \sum\limits_{i=1}^{N_P}m_i}
    \end{equation}
    where $N_P$ is the number of planetary bodies in the simulation, the subscripts 1 to $N_P$ correspond to each planetary body, and the $A$ subscript corresponds to the planetary system's central star. The second important position is $\bm{\Delta}$, the primary star's position relative to all simulated bodies' center of mass. This can be expressed \cite{kaibetal18}  as
    
    \begin{equation}
        \bm{\Delta} = \bm{X}_A-\frac{\sum\limits_{i=1}^{N_P}m_i\bm{X}_i + m_B\left(\bm{X}_B + \bm{s}\right)}{m_A + m_B + \sum\limits_{i=1}^{N_P}m_i}
    \end{equation}
    where the $B$ subscript corresponds to the binary companion of the planetary system (which again is unused here but included for completeness).

    The Plummer potential employed in our cluster environment of course introduces new potential terms into the system's Hamiltonian which are described in Ellithorpe \& Kaib\cite{Ellithorpekaib22} (see their Eq. 6).

The masses of the stars in the cluster are generated following \cite{ladalada03} and \cite{brasseretal06}, via the equation
\begin{equation}\label{eq:imf}
  M_{star} =   \frac{0.1 + (0.19 \zeta^{1.55} + 0.05 \zeta^{0.6})}{ (1.0 - \zeta)^{0.58}},
\end{equation}
where $\zeta$ is a uniformly and randomly selected number between 0 and 1. We initialized the stars positions and velocities in the cluster following the algorithm described in detail in the Appendix of Aarseth et al~\citep{aarsethetal74}, as slightly modified by Brasser et al.~\cite{brasseretal12} (see Section 3 of their work). We have performed simulations with 200, 1000, and 5000 stars.  After generating the entire  star cluster population, we randomly select one of the stars to be the host star (or to receive a binary ``companion''; see next section). Note that the selected star is not necessarily at the center of the cluster. The host-star's mass is deliberately changed to 1 solar mass. This procedure is repeated every time we create a new simulation.

In order to have a statistically robust approach, we perform a large number of simulations for each cluster configuration -- defined via the Plummer radius $R_c$ and number of stars $N_{stars}$.  We have performed at least 100 simulations for each combination of ``cluster configuration'' and `` planetary system type'', but in most cases this number is 500 or 1000. We create the cluster initial conditions  randomly selecting the masses, initial positions  and velocities of the stars (see algorithm of Aarseth et al~\citep{aarsethetal74}) for each simulation. We have performed simulations varying $R_c$ from 10~kau to 600~kau, and the number of stars varying from 200 to 5000.  Our cluster configurations probe for different cluster sizes and densities, using scenarios broadly consistent with observations and numerical models. For example, combining observational data from \cite{carpenter00} and \cite{ladalada03}, \cite{adamsetal06} suggest that the typical size of a cluster -- expressed in terms of the half-mass radius ($R_{hm}$) --  correlates with its number of stars via the following equation:
\begin{equation}
    R_{hm} =R_{sc} \left(\frac{N}{300}\right)^{\alpha},
\end{equation}
where N represents the number of stars, $R_{sc}$ is a unit scaling factor, and the power-law index $\alpha$. Previous models have considered $R_{sc}$  to range between $0.3$ and $3 {\rm~pc}$ ($\approx60{\rm~kau}$ and $1200{\rm~kau}$), $\alpha=0.5$, and $N$ to vary between $300$ to $2000$ stars \citep{proszkowadams09}. Note that the Plummer radius $R_c$ correlates with $R_{hm}$ as $R_{hm}=1.305 {\rm~R_c}$ (\citep[e.g.][]{brasseretal06}). For our simulations with $200$ stars, this would imply a half-mass radius between $0.24{\rm~pc}$ (equivalently, $R_c \approx 0.18 {\rm~pc} \approx 38 {\rm~kau}$) and $2.45{\rm~pc}$ ($R_c \approx 1.88 {\rm~pc} \approx 386 {\rm~kau}$). For our simulations with $1000$ stars, this would imply a half-mass radius between $0.54{\rm~pc}$ (equivalently, $R_c \approx 0.42 {\rm~pc} \approx 86 {\rm~kau}$) and $5.47{\rm~pc}$ ($R_c \approx 4.2{\rm~pc} \approx 865 {\rm~kau}$). 

 Other studies instead provide an initial mass-radius relation  as
\begin{equation}
    R_{hm}=0.1^{+0.07}_{-0.04}\left(\frac{M_{ecl}}{M_{\odot}}\right)^{(0.13\pm0.04)} {\rm pc},
\end{equation}
where $M_{ecl}$ is  the mass in stars of the embedded cluster \citep{markskroupa12}. In our simulations with $200$ stars, the total mass in stars is about $\sim90$$~M_{\odot}$, resulting in an initial half-radius mass varying between $0.09{\rm~pc}$ ($R_c\approx0.07{\rm~pc}\approx 14{\rm~kau}$) and $0.23{\rm~pc}$ ($R_c\approx0.18 {\rm~ pc} \approx 37{\rm~kau}$). In our simulations with $1000$ stars, the total mass in stars is $\sim450~M_{\odot}$,  resulting in a half-radius mass varying between $0.1{\rm~pc}$ ($R_c\approx0.08{\rm~pc}\approx 16.4 {\rm~kau}$) and $0.30{\rm~pc}$ ($R_c\approx0.23{\rm~pc} \approx 47.7~{\rm kau}$). Given this range of parameters, our most compact cluster scenario, with \( R_c = 10 {\rm~kau} \), represents an extreme end-member case, too compact compared to current observations.  However, our nominal cluster configurations are largely consistent with the range of scenarios considered in previous studies.

The total integration time of our nominal simulations is set to as 10~Myr, or 20 Myr. During our simulations, we kept the gas density of the cluster constant during $t_c$ years. During this phase the stars remain strongly bounded together. At $t=t_c$, the gas of the cluster instantaneously dissipates and stars are free to become field stars. We test two values for $t_{\rm c}$,  3 and 5 Myr. We have also performed a subset of simulations where the gas of the stellar cluster smoothly dissipates following an exponential decay with e-fold timescale of 0.5~Myr. In this case, we assume that the gas of the cluster completely vanishes at 5 Myr. We have found broadly equivalent results between these two scenarios.

Our cluster dissipation timescales are inspired by observations suggesting that $\gtrsim$90\% of the gas-embedded clusters disperse in  $\lesssim$10~Myr due to different star processes and feedbacks.~\cite{kroupaetal02,ballyetal98,portegiesetal10,moralesetal13,ascenso18} Only $\sim$1\% of the stellar clusters remain bounded up to $\sim$100 Myr~\cite{ladalada03,falletal09}, and they are usually refereed to as open-clusters~\cite{proszkowadams09}. For completeness purpose, and to better demonstrate the effects of the cluster for different planetary systems, we deliberately performed simulations where the cluster lifetime is also $t_{\rm c} = $50~Myr (see open symbols in Figure \ref{fig:trapping_eff}).

We have also performed two complementary sets of simulations (500 simulations each) modeling the solar system dynamical instability where, in addition to the stellar cluster perturbations (stars + Plummer potential), we added the effect of the Milk-Way (see label ``Galactic tides'' in Figure \ref{fig:trapping_eff}). We model the effects of galactic tides following previous studies\citep{kaibetal18} and we assume a galactic density $\rho_0=0.15~{\rm M}_{\odot}{\rm pc}^{-3}$.

 \subsection{Circumbinary Star Systems}   
In order to integrate the dynamics of a circumbinary planetary system  within a stellar cluster we cannot use our nominal version of the code. A  circumbinary planetary system requires to be integrated in a  different coordinate system~\cite{chambersetal02} than that used to model planetary systems around single stars. Following Kaib et al.\cite{kaibetal18}, we implemented an algorithm that builds now on the close-binary integration scheme described in Chambers et al.~\cite{chambersetal02}. We derived the new canonical coordinates and momenta  following the same method described in Kaib et al.\citep{kaibetal18,Ellithorpekaib22} and Chambers et al.\cite{chambersetal02}. In this algorithm, we solve close encounters integrating the entire system (planets, stars of the cluster, and binaries) via the Burlisch-stoer integrator. This approach may not be adequate for long-term numerical integrations (or simulations with a large number of objects) 
due to error build-up when one repetitively switches between the Sympletic and Burlisch-Stoer integration schemes~\cite{chambers99}. In order to test the performance of our algorithm and validate its use for the purposes of this work, we have performed a variety of benchmark tests. We have verified that it conserves reasonably well the Jacobi Constant associated with the circular restricted three body problem at levels of $10^{-4}-10^{-3}$.  We have also performed a suite of planet-planet scattering simulations of three giant planets orbiting close-binaries. We found that the final eccentricity distribution of planets produced in our new code and that produced in simulations using the standard Burlisch-stoer integrator are statistically indistinguishable. We have considered these tests suffice to validate the use of our new code in this study.

When modeling the effects of the  stellar cluster on circumbinary planets, the stellar cluster is initialized using  the same algorithm invoked for planets orbiting a single star (see previous section and Aarseth et al~\citep{aarsethetal74}). Our circumbinary star-systems are always composed of a ``primary'' solar-mass star (with position and velocity randomly selected from the stellar cluster population) and a ``secondary'' star with mass (in most cases but see details below)  randomly selected from the initial mass function (Eq. \ref{eq:imf}). We  have performed simulations assuming a range of initial separations and orbital eccentricities for the host binary~\cite{tokovininkiyaeva16}. Our simulations cover scenarios where the initial binary separation is  randomly selected between i)   1 and 5~au; ii)  5 and 10~au; iii) 0.2 and 1~au; and ii)  5 and 10~au with primary and secondary stars  masses set to as 1 and 0.3 ${\rm M_{\odot}}$. In all our circumbinary simulations, the binary initial orbital eccentricity is randomly selected between 0 and 0.9~. 

We have performed at least two sets  (100 simulations each) for each of the scenarios  described just above. We model dynamical instabilities in planetary systems of different orbital configurations around the binaries. We describe in the next section how we initialize our planetary systems.   We do not perform simulations for very  wide-orbit binaries (e.g. with radial separations $>$ 10-100~au). In this scenario very-wide orbit exoplanets would be probably dynamically unstable. 
\subsection{Planetary Systems}

Our simulations are designed to account for dynamical instabilities in different types of planetary systems (see labels in Figure \ref{fig:trapping_eff}). We group our simulations in 5 different categories:  i) {\it  ``Solar System: Early Instability''}, ii) ``Solar System: Ice giants formation'' , iii)``Extrasolar: Gas Giants'', iv) ``Extrasolar: Ice giants'', and v) ``Circumbinary: Gas Giants''. Note that the ``Solar System Early Instability'' is envisioned to be an dynamical instability that takes place after the accretion of the ice giants (Uranus and Neptune). Therefore, in reality, this is the final solar system giant planet instability. The term ``Early'' is invoked here only to be in line with how it is refereed to in the literature~\cite{clementetal18,raymond20,liuetal22}.

In each category of planetary systems, we consider slightly different scenarios by changing, for instance, the number of planets, planets' initial relative distances, planetary masses, initial orbits, etc. We explain now how we initialize our planetary systems.

{\it  ``Solar System: Early Instability''}
We model the solar system dynamical instability~\cite{levisonetal11,nesvorny11} assuming two different resonant configurations for the giant planets, that have been extensively studied and shown to be more successful in matching a range of solar system constraints. Our simulations fall into the 5-planets scenario~\cite{nesvorny11} (see Supplementary Figures 1 and 2). In our fiducial simulations, the giant planets start in the 3:2, 3:2, 2:1, and 3:2\cite{nesvorny11} mean motion resonances (from Jupiter, Saturn, Extra ice-giant, Uranus to Neptune). In our ``compact scenario'', the giant planets start in the 3:2, 3:2, 4:3, and 5:4 mean motion resonances\cite{gomesetal18}. In all our simulations, planets start fully formed and they evolve in a gas-free scenario. The mass of the extra ice giant, which is initially placed between Saturn and Uranus\cite{nesvorny11}, is  $\sim$10~${\rm M}_{\oplus}$. Beyond the orbit of Neptune, we added a disk of 270 equal-mass planetesimals, extending from 20 to 30~au and carrying 20~${\rm M}_{\oplus}$. The role of the planetesimal disk is to trigger the planetary instability  \citep{morbidellietal05}. Planetesimals are distributed on almost circular and coplanar orbits following a disk with surface density proportional to $r^{-1}$, where $r$ is the heliocentric distance.

{\it  ``Solar System: Ice giants formation''} In this scenario we model the migration of Earth-mass planetary embryos in a disk of gas sculpted by Jupiter and Saturn\citep{izidoroetal15c}(see section {\it Gaseous protoplanetary disk} for details on  the gas disk structure). Planetary embryos are distributed past the orbit of Saturn and are assumed to have masses between $\sim3$ and $\sim7$ ${\rm M}_{\oplus}$.  These objects start radially separated from each other by 5 to 10 mutual Hill~\cite{kokuboida00} radii.  The total mass in planetary embryos is about $\sim60$ ${\rm M}_{\oplus}$, which translates to an average of $\sim$10 planetary embryos per simulation. We model the tidal interaction of planetary embryos with their natal disk using artificial forces as described in section {\it Gas-driven migration  prescription}. In these simulations, we also added a disk of planetesimals extending from 20 to 60~au. Planetesimals interact with the gas disk via gas-drag. The main goal when adding these objects is to infer the final level of dynamical excitation that the migration and accretion of the ice giants would imprint into the primordial Kuiper-belt (see Extended Data Figure 3 and Supplementary Figures 3 and 4).

{ \it  ``Extrasolar: Gas Giants''} Our simulations modeling dynamical instabilities among gas giants are designed to match the eccentricity distribution of observations. Our  nominal giant planets are initialized following the algorithm of Beauge \& Nesvorny 2012~\cite{beuagenesvorny12}. Our simulations start with three gas giant planets in a gas-free scenario. We do not expect our results to change dramatically in a gas-disk environment because planetary dynamical instabilities during the gas disk in giant planet systems are relatively violent, particularly if they happen during the final stages of the disk\cite{legaetal14,bitschetal23}. The semi-major axis of the innermost planet is randomly selected between 1 and 5~au. Planets receive masses according to a mass generation function which allows them to be in the range of 0.5 to $\sim10 {\rm M}_{\rm jup}$ \citep{beuagenesvorny12}. We also performed simulations where the three giant planets have instead specific mass ratios. We refer to these systems as ``10J-3J-J'', ``3J-J-S'', and ``J-S-U''. The labels refer to the masses of the three planets (e.g., 3J indicates 3${\rm M_{\rm jup}}$, S indicates 1${\rm M_{\rm sat}}$, and U indicates 1${\rm M_{\rm ura}}$ ).  These simulations are unconstrained by observations and serve as a complementary exploration of parameters. HR8799-like systems have initially four instead of three planets, and planetary masses are randomly selected between 0.5 and 2 times the  estimated masses of the HR8799 planets\citep{wangetal18} (6.7, 5.4, 6.2, and 6.6~${\rm M}_{\rm jup}$). Also in HR8799-like system simulations, the innermost planet is initially placed between 10 and 20~au. Subsequent planets are placed in 2:1 mean motion resonance with the previous adjacent planet.  Our simulations of HR 8799-like systems are purely exploratory and are not constrained by observational data.

{ \it  ``Extrasolar: Ice Giants''} We model dynamical instabilities in systems of ice giant planets considering 6 planets in the system. In these simulations, the gaseous protoplaentary disk is assumed to have already dissipated.  The orbital radius of the innermost planet is randomly selected between 2 and 10~au, and subsequent planets are placed at either 2:1 or 3:2 mean motion resonances with adjacent planets. The initial individual mass of planetary embryos is set to as 20~${\rm M}_{\oplus}$.

 { \it  ``Circumbinary: Gas Giants''} We initialize planets orbiting the binary following the same recipe\citep{beuagenesvorny12} used for our ``Extrasolar: Gas giants'' systems.   In this particular case, however, we performed simulations not only with three giant planets but also scenarios with only two gas giants. In all our simulations planetary masses are sampled between  1 and $\sim$10 ${\rm M}_{{\rm jup}}$. Planets are initially placed on orbits with semi-major axis always larger than the semi-major axis of the system binary (see previous section for details on binary configurations explored in this work). Planets and binaries are initialized with orbital inclinations randomly selected between $\sim$0.01 and $\sim$0.1 degrees, and other orbital angles are randomly selected between 0 and 360 degrees. Planets are initially placed on virtually circular orbits.
 
\subsection{Gaseous protoplanetary disk}

As explained before, in our simulations modeling the dynamical evolution of giant exoplanets (planet-planet scattering simulations),  the Solar System dynamical instability, and the dynamical evolution of circumbinaries planets, we assume for simplicity that the  star (or binary) natal disk has already dissipated. In our simulations of the accretion of ice giants, in additional to the cluster effects,  we model the gas disk around the young Sun considering an underlying 1D gaseous protoplantary disk obtained from 2D locally isothermal hydrodynamical simulations\citep{morbidellicrida07}. We used the same gas disk profile used in  previous simulations\citep{izidoroetal15a}. The disk initial surface density corresponds to the traditional minimum mass solar nebula model \citep{hayashi81}. In the hydrodynamical simulations, Jupiter and Saturn are considered to be on fixed orbits and near the 3:2 mean motion resonance. Jupiter is at 5.25~AU and Saturn is at about 7.18~AU. The simulations evolves until the planets have opened up gaps in the disk and the gas disk has reached an equilibrium state.  To generate the 1D disk profile from the output of the 2D hydrodynamical simulations we averaged the final gas disk profile  over the azimuthal direction. We use the  1D disk profile as input for our N-body code as\cite{izidoroetal15c}. In this work we use a single gas disk profile, but we test the impact of different gas surface density in our simulations. We rescale our original gas disk profile by a factor $f_{gas}=0.5$ or 0.7 in order to account for initially lower-mass disks. In all our simulations of the accretion of Uranus and Neptune, the host star is assumed to be fully formed and to be solar mass.

The disk aspect ratio in our simulations is modeled via the following equation
\begin{equation}
{\rm h(r)=\frac{H(r)}{r}=0.033r^{0.25},}
\label{eq:aspect}
\end{equation}
where $H$ is the disk scale height. We also use the standard alpha-viscosity approach \citep{shakurasunyaev73} and set $\alpha = 0.002$. Below we describe how we model the gravitational interaction of planetary bodies with the gas disk and also gas-drag on planetesimals.

\subsection{Gas-driven migration prescription}

In our simulations modeling the accretion of the ice giants we account for gas-driven planetary migration. Massive protoplanetary embryos interact gravitationally with the gas disk and migrate. We also follow Izidoro et al.\cite{izidoroetal15c} to model the gas-driven migration of protoplanetary embryos. 

The total gas disk torque on a massive protoplanetary embryo is given by 
\begin{equation}
{\rm \Gamma_{tot} = \Gamma_L\Delta_{L} + \Gamma_C\Delta_{C}},
\label{eq:totaltorque}
\end{equation}
where $\Delta_{L}$ and $\Delta_{L}$ corresponds to the contributions from the Lindblad and co-orbital torques for the case where the planets have circular orbits. One shall modify these expressions by the factors   $\Gamma_L$ and $\Delta_{C}$ to account for the  reductions on the respective torques when the planet evolves on eccentric or inclined orbits. $\Delta_{L}$ is given as
\begin{equation}
{\rm \Delta_L = \left[   P_e + \frac{P_e}{|P_e|} \times \left\lbrace 0.07 \left( \frac{i}{h}\right)  + 0.085\left( \frac{i}{h}\right)^4 -  0.08\left( \frac{e}{h} \right) \left( \frac{i}{h} \right)^2 \right \rbrace \right] ^{-1} } ,
\end{equation}
where
\begin{equation}
{\rm P_e = \frac{1+\left( \frac{e}{2.25h}\right)^{1.2} +\left( \frac{e}{2.84h}\right)^6}{1-\left( \frac{e}{2.02h}\right)^4}}.
\end{equation}

The Lindblad torque is set as
\begin{equation}
{\rm \Gamma_L= (-2.5 -1.5\beta + 0.1x)\Gamma_0},
\end{equation}
where $x$ and $\beta$ are the negative of the local (at the location of the planet) gas surface density and temperature gradients, respectively.

Following  previous studies\cite{cresswellnelson08,colemannelson14}, ${\rm \Delta_{C}}$ is written as
\begin{equation}
{\rm \Delta_{C}=exp\left(\frac{e}{e_f} \right)\left\lbrace 1-tanh\left(\frac{i}{h} \right)\right\rbrace   },
\end{equation}
 ${\rm e_f}$ is defined as
\begin{equation}
{\rm e_f = 0.5h + 0.01},
\end{equation}
where i and e are the planet's eccentricity and orbital inclination, respectively. In the locally isothermal limit, the co-orbital torque is given by 
\begin{equation}
{\rm \Gamma_C=\Gamma_{hs,baro}F(p_{\nu})G(p_{\nu}) + (1 - K(p_{\nu}))\Gamma_{c,lin,baro} + (1 - K(p_{\nu}))\Gamma_{c,lin,ent}}.
\end{equation}

The barotropic and entropy-related components of the co-orbital torque are given by ${\rm \Gamma_{hs,baro}}$, ${\rm \Gamma_{c,lin,baro}}$, and ${\rm \Gamma_{c,lin,ent}}$ as
\begin{equation}
{\rm \Gamma_{hs,baro}= 1.1\left( \frac{3}{2}-x\right) \Gamma_0},
\end{equation}
\begin{equation}
{\rm \Gamma_{c,lin,baro}= 0.7\left( \frac{3}{2}-x\right) \Gamma_0},
\end{equation}
and
\begin{equation}
{\rm \Gamma_{c,lin,ent}= 0.8\beta \Gamma_0,}
\end{equation}

The scaling torque is given by ${\rm \Gamma_0=(q/h)^2\Sigma_{gas} r^4 \Omega_k^2}$ and it is calculate at the planet's location. The planet-star mass ratio is denoted by $q$, h is the gas disk aspect ratio as described in Eq. \ref{eq:aspect}. ${\rm \Omega_k}$ represents the  planet's orbital Keplerian frequency and ${\rm \Sigma_{gas}}$ is the gas surface density at the planet's location. 

To account for torque saturation due to the disk viscous and thermal properties of the disk we follow Paardekooper et al.\cite{paardekooperetal10,paardekooperetal11} and invoked the following equations 
\begin{equation}
{\rm p_{\nu} = \frac{2}{3}\sqrt{\frac{r^2\Omega_k}{2\pi\nu}x_s^3}},
\end{equation}
where ${\rm x_s}$ represents the half-width of the horseshoe region and is given gy
\begin{equation}
{\rm x_s=\frac{1.1}{\gamma^{1/4}}\sqrt{\frac{q}{h}}=1.1\sqrt{\frac{q}{h}. }}
\end{equation}

The migration timescale is calculated as
\begin{equation}
{\rm t_m =- \frac{L}{\Gamma_{tot}}},
\end{equation}
where $L$ is the planet's orbital angular momentum and $\Gamma_{tot}$ is the total torque given by Eq. \ref{eq:totaltorque}.

The timescales for damping of orbital inclination and eccentricity are also modeled as in \cite{cresswellnelson08} using the following equations:
\begin{equation}
{\rm t_e = \frac{t_{wave}}{0.780} \left(1-0.14\left(\frac{e}{h/r}\right)^2 + 0.06\left(\frac{e}{h/r}\right)^3    + 0.18\left(\frac{e}{h/r}\right)\left(\frac{i}{h/r}\right)^2\right),}
\end{equation}
and
\begin{equation}
{\rm t_i = \frac{t_{wave}}{0.544} \left(1-0.3\left(\frac{i}{h/r}\right)^2 + 0.24\left(\frac{i}{h/r}\right)^3    + 0.14\left(\frac{e}{h/r}\right)^2\left(\frac{i}{h/r}\right)\right),}
\end{equation}
where
\begin{equation}
{\rm t_{wave} = \left(\frac{M_{\odot}}{m_p}\right)  \left(\frac{M_{\odot}}{\Sigma_{gas} a^2}\right)\left(\frac{h}{r}\right)^4 \Omega_k^{-1}}
.\end{equation}
The quantities ${\rm M_{\odot}}$, ${\rm a_p}$, ${\rm m_p}$, ${\rm i}$, ${\rm  and e}$ represent the sun's mass, the planet's orbital semi-major axis and mass, orbital inclination, and eccentricity, respectively.  Note that these timescales are particularly derived from fits to hydrodynamical simulations for planets with eccentricities of about $\lesssim0.3-0.5$, and some of our planets acquire eccentricities higher than those during the gas disk phase. Equivalent prescriptions have been used in previous studies \cite{colemannelson14,bitschetal23} but different prescriptions also exist \cite{idaetal20}.

In our code, artificial accelerations are included to mimic the effects of planet-disk tidal interactions following previous studies\citep{papaloizoularwood00}

\begin{equation}
{\rm \mathbf{a}_m = -\frac{\mathbf{v}}{t_m}}
\end{equation}

\begin{equation}
{\rm \mathbf{a}_e = -2\frac{(\mathbf{v.r})\mathbf{r}}{r^2 t_e}}
\end{equation}

\begin{equation}
{\rm \mathbf{a}_i = -\frac{v_z}{t_i}\mathbf{k},}
\end{equation}
where ${\rm \mathbf{k}}$ is the vertical component of the reference frame. In our simulations

\subsection{Gas drag}

In our simulations modelling the accretion of Uranus and Neptune via giant impacts among Earth-mass planetary embryos, we also assume the presence of a disk of planetesimals beyond 20~au. This disk is envisioned to be the primordial Kuiper belt, which is sculpted during the accretion of Uranus and Neptune~\cite{ribeiro20} (and subsequently when the giant planets evolve to their current orbits\cite{nesvorny11}).  We model the effects of gas drag on planetesimals by adding the following acceleration when solving for their motion

\begin{equation}
{\rm {\mathbf a}_{drag} =- \frac{3C_d \rho_g v_{rel} \mathbf{v}_{rel}}{8 \rho_p R_p}},
\label{eq:drag}
\end{equation}
where ${\rm v_{rel}}$ is the planetesimal's relative velocity to the surrounding gas. In Eq. \ref{eq:drag}, ${\rm C_d}$ denotes the drag coefficient,  ${\rm \rho_p}$ is the planetesimal's bulk density, and ${\rm R_p}$ is the planetesimals physical radius. The gas density is calculated as  
\begin{equation}
\rho_{gas} = \frac{\Sigma_{gas}}{2\sqrt{\pi}}exp(-z^2/H^2),
\end{equation}
where $z$ is the planetesimal's z-component. To compute the gas drag coefficient we follow \citep{brasseretal07}. Planetesimals in the primordial Kuiper-belt disk are assumed to have diameters of 100~km, a representative-size of planetesimals formed by streaming instability~\cite{johansenetal07,simonetal16}. Our simulations are computationally intensive so performing simulations to test the effects of different planetesimals size would be impractical and is beyond the scope of our work.

\vskip .2in

\noindent{\bf Data availability}\\
Simulation data that support the findings of this study or were used to make plots are available from the corresponding author upon reasonable request. The source data of the main figures of the paper are available at \href{https://andreizidoro.com/simulation-data}{this link}.

\noindent{\bf Code availability}\\ Simulations  presented here were performed using modified versions of the {\tt Mercury} N-body integrator\cite{chambers99}, publicly available at \href{https://github.com/4xxi/mercury}{this link}.

\begin{addendum}
 \item [Acknowledgments]
A. Izidoro is grateful to Rajdeep Dasgupta for insightful discussions, help with proofreading, valuable input on manuscript clarity, and partial financial support for this project. A.~Izidoro. and N.~A.~K thank support from the NASA Emerging Worlds Program Grant 80NSSC23K0868. Contributions from N.~A.~K were also supported by NASA Exoplanets Research Program grant 80NSSC19K0445 and NSF CAREER Award 2405121. S.N.R acknowledges funding from the Programme Nationale de Planetologie (PNP) of the INSU (CNRS), and in the framework of the Investments for the Future programme IdEx, Université de Bordeaux/RRI ORIGINS. A.~M. acknowledges support from ERC grant 101019380 HolyEarth. This work was supported in part by the Big-Data Private-Cloud Research Cyberinfrastructure MRI-award funded by NSF under grant CNS-1338099 and by Rice University's Center for Research Computing (CRC). 
\item [Author Contributions Statement] A.~Izidoro, S.~N.~R., N.~A.~K, and A.~M.  conceived the original idea of this project. N.~A.~K. and A.~Izidoro wrote and tested the adapted versions of MERCURY code used in this work. A.~Izidoro performed  numerical simulations, analyzed results and prepared all figures of the paper. S.~N.~R., A.~Izidoro, and N.~A.~K drafted the manuscript with inputs from A.~M., and A.~Isella. All authors contributed to the interpretation and discussion of the results, writing and editing of the  manuscript.
\item[Competing Interests Statement] The authors declare no competing interests.
\item[Correspondence] Correspondence and requests for materials should be addressed to A. Izidoro  (izidoro [at] rice.edu)
\end{addendum}



\beginextendeddata

\begin{figure}[ht]
\centering
\includegraphics[scale=0.7]{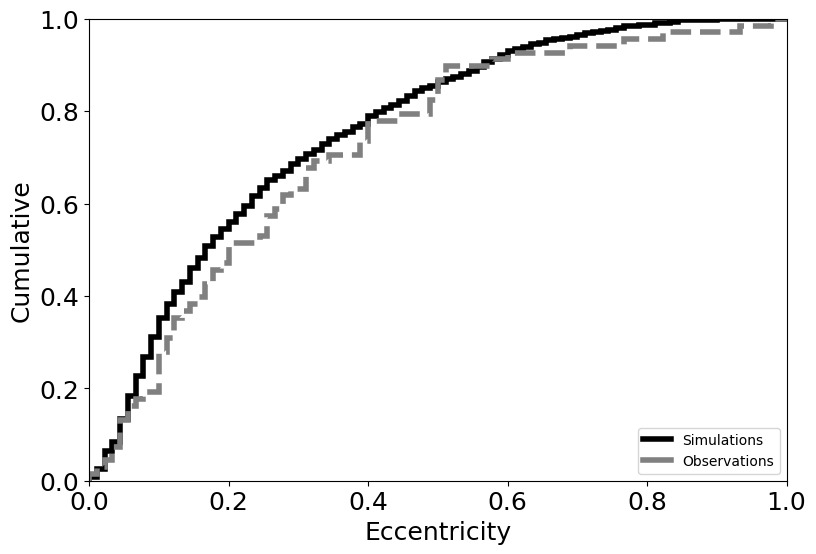}
    \caption{Final eccentricity distribution of giant exoplanets surviving in the inner systems after dynamical instabilities. We show only giant planets with semi-major axis smaller than 40~au. he eccentricity distribution of  radial velocity exoplanets are shown in grey.}
    \label{fig:eccentricitydistribution}
\end{figure}

\begin{figure}[ht]
\centering
\includegraphics[scale=0.24]{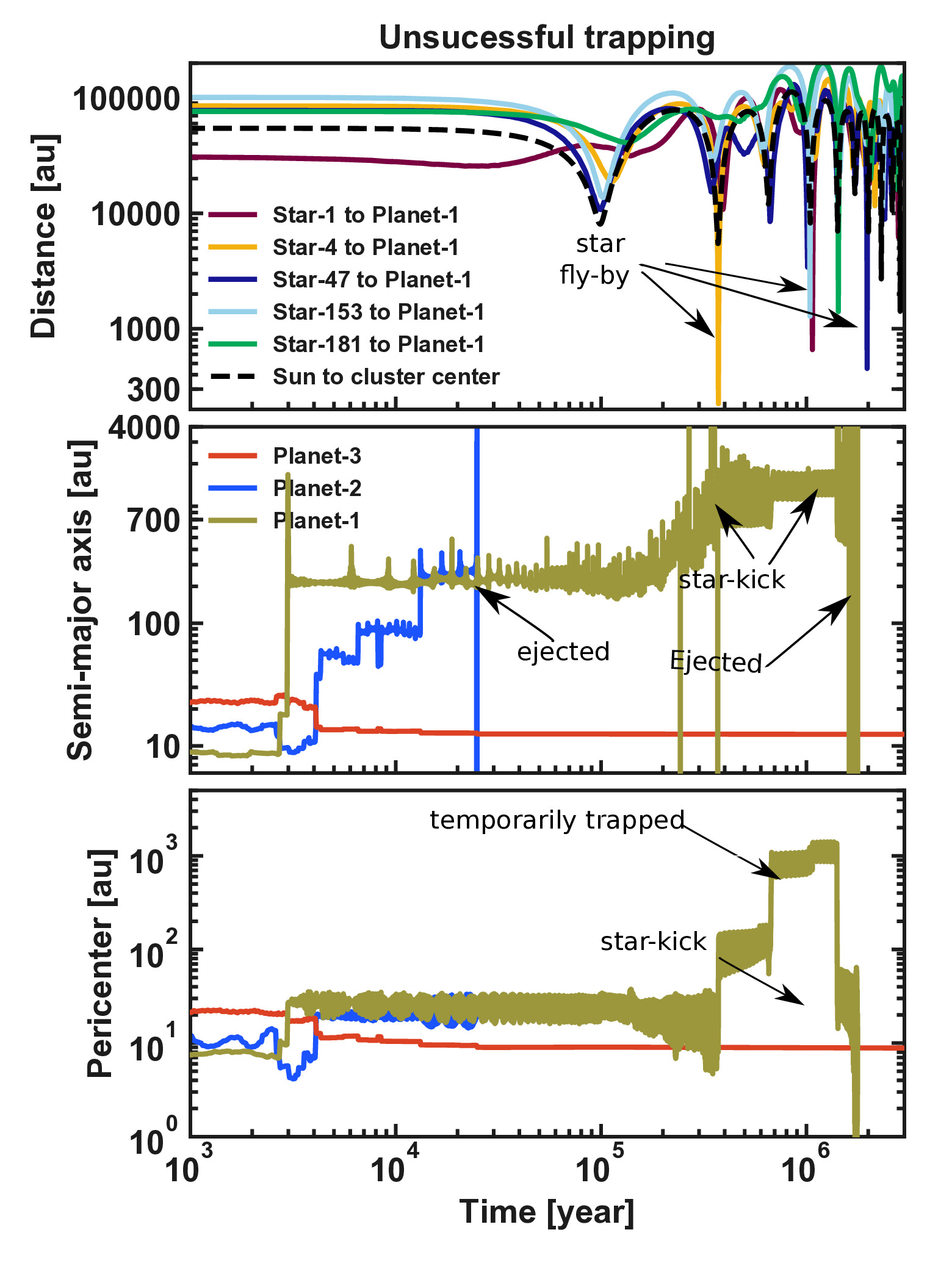}
    \caption{Unsuccessful trapping of a wide-orbit planet in an exoplanet gas giant instability simulation. A planetary dynamical instability takes place at about 3 kyr which scatters Planet-1 and 2 on wide orbits. Planet-2 is ejected from the host star at ~25 kyr. Planet-1 evolves onto a high eccentricity orbit and is eventually kicked by Star-4 at about 0.4 Myr. A passage of the host star near the cluster barycenter at 0.7 Myr also affects the orbit of Planet-1. Planet-1 is ultimately ejected from the host start due to multiple flybys of stars 1, 47, 153, and 181, which take place after 1-2 Myr. The simulation stops at 3 Myr. Orbital elements are given with respect to the barycenter of the host-star system}
    \label{fig:failedtrapping}
\end{figure}

\begin{figure*}[ht]
\vspace{-1.cm}
\centering
\includegraphics[width=.68\linewidth]{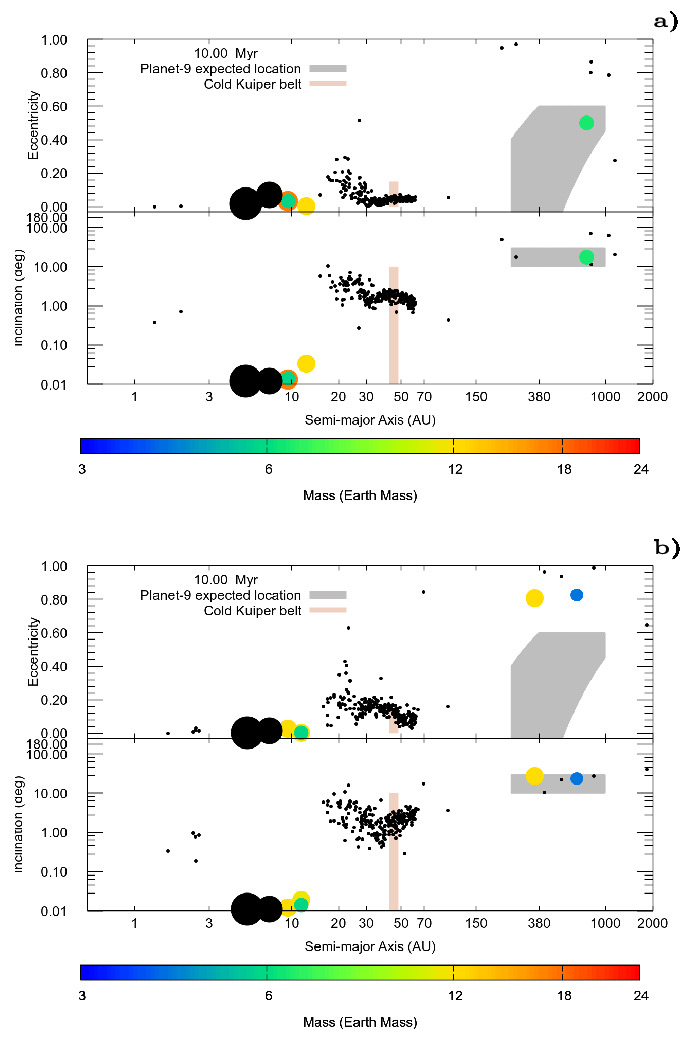}
\caption{\footnotesize Planetary system architecture at the end of the gas disk phase in two different simulations of the accretion of Uranus and Neptune that  broadly match solar system constraints (almost unitary mass ratio of the ice giants, two planets with masses larger than about $12~{\rm M}_{\oplus}$, and a dynamically cold planetesimal disk population consistent with the dynamically cold kernel of the Kuiper-belt). Each plot is composed by two stacked panels. The top-component shows semi-major axis versus eccentricity.  The bottom-component shows semi-major axis versus orbital inclination. Jupiter and Saturn are represented by the big black filled circles showed at about  5.25 and about 7.18 AU (near the 3:2 MMR). Color-coded circles represent the formed ice giants and their sizes scale as $M^{1/3}$, where M is the mass. The color-coding  shows the masses of the final planets. The small black circles represent primordial planetesimals. The gray-region show the expected location of Planet-9. The orange-ish regions ($a\approx45$~au, $e<0.1$, and $i<10$ degrees) are used to represent the Kernel of the current Kuiper-belt. Note that, in both simulations, the   dynamical excitation of the primordial disk -- after the accretion of the ice giants -- is broadly consistent with the cold dynamical architecture of the Kuiper-belt kernel.}
\label{fig:UN-P9-example}
\end{figure*}

\clearpage
\noindent{\bfseries References}\setlength{\parskip}{12pt}%

\bibliography{mybib}

\clearpage




\beginsupplement

This Supplementary Information presents additional simulation results that support and extend the findings discussed in the main text. It includes simulations not shown in the main paper, as well as supplementary figures and movies that provide further context and detail.

\tableofcontents
\newpage

\section{Formation of the Oort Cloud in Early Solar System Instability Simulations}

In this work, we do not follow the long-term dynamical evolution of the solar system ($\sim$4.5 Byr) in order to fully model the formation and evolution of the Oort Cloud. However, our results suggest that an early dynamical instability during the cluster embedded phase is probably consistent with the formation of the Oort cloud. 

We have selected a few examples of solar system dynamical instabilities that produced good solar system analogues and also wide-orbit exoplanets. For comparison purposes, we also show the results of two solar system dynamical instabilities where we neglect the effects of the stellar cluster. Selected cases are broadly representative of our simulations.

We recall that our simulations modeling the solar system dynamical instability assume the presence of a primordial disk of planetesimals  beyond the orbit of Neptune (extending from $\sim$21 to 30 au) initially carrying about $\sim$20~${\rm M}_{\oplus}$~\citep{malhortra1995,fernandezip84,gomesetal05,nesvornymorbidelli12,batyginetal12,deiennoetal18,clementetal18}. During the onset of the instability, the giant planets gravitationally interact with the planetesimal primordial disk. Consequently, planetesimals are scattered on very eccentric orbits creating the so-called ``scattered'' Kuiper-belt disk and populating the Oort cloud~\citep{malhortra1995,nesvorny18}.

If the entire primordial disk population is scattered during the phase when the Sun is still embedded in the cluster is reasonable to expect that  (the outer part of) the Oort cloud may not form at all, or may be too anemic and inconsistent with the current population of long-period comets~\citep{kaibquinn08,nordlanderetal17,brassermorbidelli13}. Stars of the cluster may prevent the formation of the Oort cloud by strongly perturbing and unbinding (from the Sun) planetesimals reaching orbits with semi-major axes between  $\sim10,000$~au and $\sim100,000$~au\citep{nesvornyetal17}. In order to check whether this is an issue or not affecting the early dynamical instability scenario, we compute the fraction  of bound planetesimals beyond Jupiter (q$>$5~au; a$<$100~kau), the fraction of planetesimal in the primordial Kuiper-belt disk (a$<$50~au; e$<$1), the fraction of unbound planetesimals (e$>$1), the fraction of planetesimal at q$>$150~au, a$<$~10~kau, and e$<$0.95 (hereafter defined as the inner-edge of the Oort cloud), and the fraction of planetesimals at a$>$10~kau, a$<$~100~kau, and e$<$0.95 (hereafter defined as the Oort cloud). Fractions are computed relative to the initial number of planetesimals in the disk.

\begin{figure}[!h]
\centering
\includegraphics[scale=1.3]{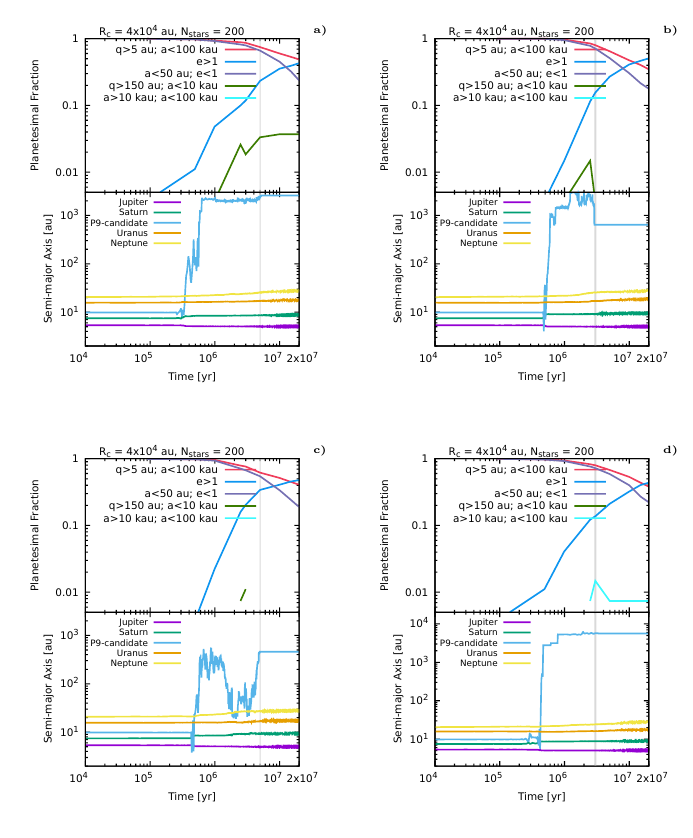}
    \caption{Four solar system early dynamical instability simulations producing reasonable solar system analogues with trapped wide-orbit planets. Each sub-figure contains two panels. The top panel shows the fractions of: i)  planetesimals bound to the Sun (red); ii)  planetesimals ejected from the Solar System (blue); ii) planetesimals residing in the Kuiper-belt region (purple); iii) planetesimals near the inner edge of the Oort cloud (green; e$<$0.95), and iv) planetesimals in the Oort cloud (magenta;  e$<$0.95). The bottom panel shows the evolution of the giant planets semi-major axes. The grey vertical bar shows the time of the cluster dissipation ($t_c$).}
    \label{fig:Oort}
\end{figure}

Supplementary Figure \ref{fig:Oort} shows  selected solar system dynamical instabilities that produced wide-orbit exoplanets. The configuration of the stellar cluster in each simulation is given at the top of its respective plot. The top-panel of each plot shows the fraction of  planetesimals in different regions and dynamical states. The bottom-panel shows the evolution of the planets. The thin-vertical line in each panel marks the time of the cluster dispersal.


Supplementary Figure \ref{fig:Oort} shows that at the end of our simulations -- when the cluster has already significantly dispersed -- about $\sim$40-50\% of the original planetesimal population is still bound to the Sun, and about $\sim$20\% is still in the primordial disk (see red and purple lines). Our simulations suggest that the fraction of planetesimals trapped near the inner edge of the Oort cloud can reach up to $\sim$4\% of the disk population (green line on the top-left panel of Figure \ref{fig:Oort}).  However, depending on the Sun evolution in the cluster, this fraction can be as low as $\sim$0\% (green lines on the right panels of Figure \ref{fig:Oort}). The fraction of planetesimals trapped in the Oort cloud during 20~Myr (blue line) is systematically lower than 0.5-1\%, and in most cases it is virtually zero. Note that these values should not be interpreted as final trapped fractions. In order to properly simulate the formation of the Oort cloud we would need to include the effects of the galactic tide, passing stars, and perform long-term numerical integrations for the age of the Solar System. This is beyond the scope of this work. Our intention here is only to suggest -- using order of magnitude arguments  -- that our early solar system instability simulations are broadly consistent with the formation of the Oort cloud. 

Following this reasoning, in a scenario where the primordial disk beyond Neptune carries initially $\sim$20-40~${\rm M}_{\oplus}$ in planetesimals, and invoking that the overall trapping efficiency of planetesimal in the Oort Cloud is about 10\%  when including the effects of galactic tides and passing stars \cite{brasseretal06,kaibquinn08,brassermorbidelli13}, our simulations would be able to produce an Oort Cloud carrying about $\sim$0.8-2~${\rm M}_{\oplus}$ in planetesimals (accounting only for planetesimals bound at 20~Myr). This range is broadly consistent with  current estimates of the total mass in the Oort cloud~\citep{donesetal15}.

In Supplementary Figure \ref{fig:Oort_b}, we show the evolution of planetesimal populations in simulations where we neglect the effects of the stellar cluster. Results of Supplementary Figures \ref{fig:Oort} and \ref{fig:Oort_b} are broadly equivalent. This is interesting because it shows that the embedded cluster phase does not necessarily enhance the early ejection ($<$20~Myr) of planetesimals from the Solar System. One difference between these two sets of simulations, however,  is that in the absence of perturbations from the cluster the fraction of trapped planetesimal near the inner edge of the Oort cloud is always virtually zero. In contrast, it may yield up to a few percent when the stellar cluster is considered. In this work, we are not able to constrain the population ratio between the inner and outer Oort cloud\cite{nesvornyetal17}  but this remains as an interesting avenue for future studies. 

Altogether, this analysis suggests that an early instability during the embedded cluster phase is broadly consistent with the Oort cloud formation if about $\sim$10\% of planetesimal population bound to the Sun at 20~Myr is trapped in the Oort cloud during the subsequent 4.5~Gyr evolution of the Solar System\cite{brasseretal06,kaibquinn08,brassermorbidelli13}. It remains as a noble exercise, however,  to further investigate this issue using high resolution and long-term numerical simulations, particularly in the presence of  Planet-9\citep{nesvornyetal17}.





\begin{figure}[!h]
\centering
\includegraphics[scale=1.3]{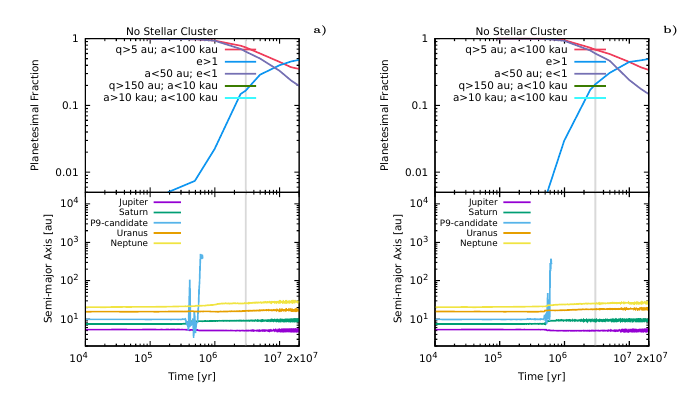}
    \caption{As Supplementary Figure \ref{fig:Oort} but for two early solar system dynamical instability simulations where the effects of the stellar cluster were neglected. As expected,  wide orbits planets are not trapped in these cases, they are instead ejected. }
    \label{fig:Oort_b}
\end{figure}

\newpage

\section{Example of final systems produced in simulations modeling the accretion of the ice giants}\label{sec:results}

\begin{figure*}[!h]
\vspace{-1cm}
\centering
\includegraphics[width=.7\linewidth]{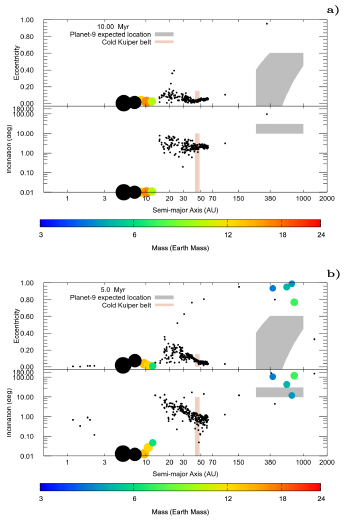}

\caption{Planetary system architecture at the end of the gas disk phase in two simulations of the accretion of Uranus and Neptune that  broadly match solar system constraints. The top-component  shows semi-major axis versus eccentricity.  The bottom-component shows semi-major axis versus orbital inclination. Jupiter and Saturn are represented by the big black filled circles showed at about  5.25 and about 7.18 AU (near the 3:2 MMR). Color-coded circles represent the formed ice giants and their sizes scale as $M^{1/3}$, where M is the mass. The color-coding  shows the masses of the final planets. The small black circles are  planetesimals, representing the primordial Kuiper-belt. The gray-region shows the expected location of Planet-9. The orange-ish regions (a$\approx$45~au, e$<$0.1, and i$<$10 degrees) are used to represent the Kernel of the current Kuiper-belt.}
\label{fig:UN-P9-example-sup-1}
\end{figure*}

\begin{figure*}[!h]
\centering
\vspace{-1cm}

\includegraphics[width=.65\linewidth]{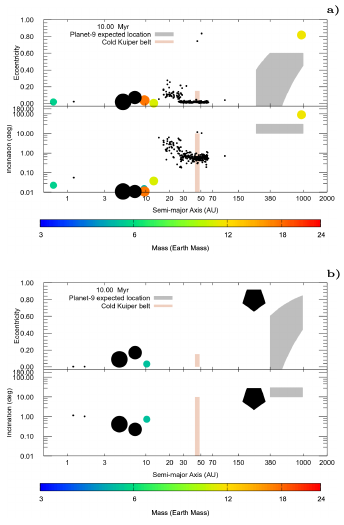}

\caption{Planetary system architecture at the end of the gas disk phase in two simulations of the accretion of Uranus and Neptune that failed in matching the solar system. Each plot is composed by two stacked panels. The top-component shows semi-major axis versus eccentricity.  The bottom-component shows semi-major axis versus orbital inclination. Jupiter and Saturn are represented by the big black filled circles showed at about  5.25 and about 7.18 AU (near the 3:2 MMR). Color-coded circles represent the formed ice giants and their sizes scale as $M^{1/3}$, where M is the mass. The color-coding  shows the masses of the final planets. The small black circles are  planetesimals, representing the primordial Kuiper belt. The gray-region shows the expected location of Planet-9. The orange-ish regions (a$\approx$45~au, e$<$0.1, and i$<$10 degrees) are used to represent the Kernel of the current Kuiper-belt. The pentagon represents a cluster-star -- that during a close-encounter with the Sun  -- ejects all objects between 20 and 60~au. }
\label{fig:UN-P9-example-sup-2}
\end{figure*}

In this section we show additional examples of simulations that produced ``good'' and ``bad'' solar system analogues. Supplementary Figure  \ref{fig:UN-P9-example-sup-1}-top panel shows a good solar system analogue with no trapped wide-orbit planet. Supplementary Figure  \ref{fig:UN-P9-example-sup-1}-bottom panel shows another extreme scenario where 4 wide-orbit planets are trapped. Three of these planets have high eccentricities, so they are not necessarily long-term stable. Supplementary Figure \ref{fig:UN-P9-example-sup-2}-top panel shows an example where a wide-orbit planet is trapped but an 5~${\rm M_{\oplus}}$ is scattered into the inner solar system, which is not consistent with our current system. Finally,  Supplementary Figure \ref{fig:UN-P9-example-sup-2}-bottom panel shows a scenario where the solar system is destroyed due to a close-encounter with a star. 

In about 50\% of our simulations modeling the accretion of Uranus and Neptune, the planetesimal disk representing the primordial Kuiper belt remains dynamically cold at levels consistent with the Kuiper-belt kernel population. On average, our simulations produce jumper-planets (embryos scattered into the inner solar system) in about 50\% of the cases. We formed good solar system ice giants analogues (ice-giants mass ratio between 1 and 1.5, masses larger than 12$M_{\oplus}$, and each ice giant experiencing at least one impact) in about $\sim$5-10\% of the simulations. Our simulations trap wide-orbit planets at the  very expected location of Planet-9\citep{batygin19} in about $\lesssim1$\% of our simulations modeling the accretion of Uranus and Neptune.


\newpage

\section{Dynamical instabilities in HR8799-like systems}

In most of our simulations, planets are initially located inside $\sim$20-40~au. Consequently, dynamical instabilities alone are typically not capable of producing planets with semi-major axes larger than $\sim$100~au and orbital eccentricities lower than $\sim$0.8-0.9 (that would look like our wide-orbit trapped planets). However, in simulations modeling instabilities of HR-8799-like systems, planets are very massive and initially have very large orbital separations (planets are between  $\gtrsim$10 and $\gtrsim$70~au). In this scenario, dynamical instabilities are  strong enough to scatter planets to larger semi-major axes  ($\sim$300-400~au), yet low-eccentricity orbits. 

\begin{figure*}[!h]
\centering
\includegraphics[width=0.7\linewidth]{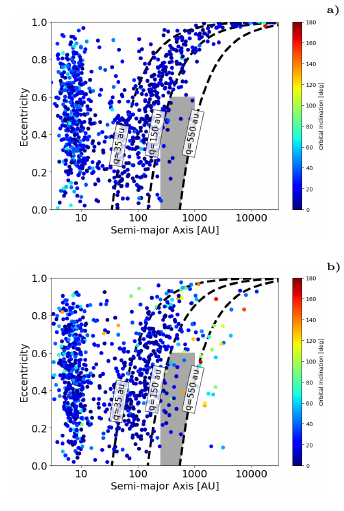}
\caption{Final distribution of planets in instability simulations of HR8799-like planetary systems. Simulations start with four super-Jupiter mass planets (see Methods). The top-panel shows the results of 500 simulations neglecting the effects of  stellar clusters. The bottom-panel shows the results of 500 simulations with planetary systems embedded in stellar clusters. In all simulations of the bottom-panel the clusters contains 200 stars and a Plummer radius of 40,000~au.  Dashed lines indicate regions where the pericenters are 35, 150, and 550 AU. The gray regions represent the predicted location of Planet 9 and are included solely for reference and comparison.}
\label{fig:HR8799-example}
\end{figure*}

Supplementary Figure \ref{fig:HR8799-example}-top panel shows the final distribution of planets produced in one set of 500 simulations of HR8799-like systems where we neglect the effects of the stellar cluster. As can be seen, a few planets overlap with the gray region that covers semi-major axes between $\sim$400 and 1000~au and orbital eccentricities lower than 0.8 (some of them may not be dynamically decoupled from the inner system yet).  These specific planets were produced by pure planetary instabilities, without kicks of the stellar cluster. Supplementary Figure \ref{fig:HR8799-example}-bottom panel shows the outcomes of instabilities in the same systems of the top-panel, but when we account for the effects of the cluster. In addition to wide-orbit planets produced via pure planetary instabilities, a significant number of planets is also trapped at low-eccentricity and wide-orbits due to the effects of the stellar cluster (many at $a>$1000~au). A clear difference between these two scenarios is seen in the orbital inclination of their respective wide-orbit planet populations.  Wide-orbit planets produced via pure planetary instabilities usually show orbital inclinations lower than 40-60 degrees. Wide-orbit planets trapped due to the effects of the stellar cluster tend to show a range of orbital inclinations -- varying from virtually 0 to 180 degrees -- because their orbital planes are tilted by stars' flybies (or kicks from the gas cluster) with no preferential direction.

\section{Gaseous disk Model} 

\begin{figure}
\centering
\includegraphics[scale=1.2]{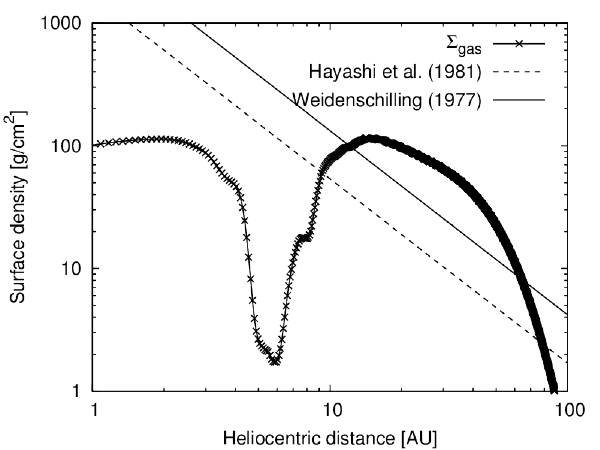}
\caption{One-dimensional gas surface density profile produced via azimuthal averaging of a 2D disk from hydrodynamical simulations. Jupiter and Saturn are considered on fixed orbits and reside at 5.25~AU and 7.18~AU, respectively. We also plot the gas surface density profiles of the minimum mass solar nebula models for reference \citep{weidenschilling77,hayashi81}. }
\label{fig:gasdensity}
\end{figure}

Figure \ref{fig:gasdensity} shows the gas disk surface density profile used in our simulations modelling the accretion of Uranus and Neptune. 


\end{document}